
\def\oneone{\rlap 1\mkern4mu{\rm I}}
\overfullrule=0pt
\def\singlespace{\normalbaselines}
\def\oneandahalfspace{\baselineskip=16pt plus 1pt
\lineskip=2pt\lineskiplimit=1pt}

\def\np{\vfill\eject}
\def\nl{\hfil\break}

\def\nofirstpagenoten{\nopagenumbers\footline={\ifnum\pageno>1
\tenrm\hss\folio\hss\fi}}
\def\nofirstpagenotwelve{\nopagenumbers\footline={\ifnum\pageno>1
\twelverm \hss\folio\hss\fi}}
\def\leaderfill{\leaders\hbox to 1em{\hss.\hss}\hfill}


\parindent=20pt
\def\narrow{\advance\leftskip by 40pt \advance\rightskip by 40pt}
\baselineskip=20pt
\def\AB{\bigskip
        \centerline{\bf ABSTRACT}\medskip\narrow}
\def\nonarrower{\advance\leftskip by -40pt\advance\rightskip by -40pt}
\def\AE{\bigskip\nonarrower}

\def\boxit#1{\vbox{\hrule\hbox{\vrule\kern3pt
        \vbox{\kern3pt#1\kern3pt}\kern3pt\vrule}\hrule}}

\def\gtorder{\mathrel{\raise.3ex\hbox{$>$}\mkern-14mu
             \lower0.6ex\hbox{$\sim$}}}
\def\ltorder{\mathrel{\raise.3ex\hbox{$<$}|mkern-14mu
             \lower0.6ex\hbox{\sim$}}}
\def\dalemb#1#2{{\vbox{\hrule height .#2pt
        \hbox{\vrule width.#2pt height#1pt \kern#1pt
                \vrule width.#2pt}
        \hrule height.#2pt}}}

\font\fourteentt=cmtt12 scaled \magstep1\font\fourteenbf=cmbx12
scaled \magstep1
\font\fourteenrm=cmr12 scaled \magstep1\font\fourteeni=
cmmi12 scaled \magstep1
\font\fourteenss=cmss12 scaled \magstep1
\font\fourteensy=cmsy10 scaled \magstep2 \font\fourteensl=
cmsl12 scaled \magstep1
\font\fourteenex=cmex10 scaled \magstep2 \font\fourteenit=
cmti12 scaled \magstep1
\font\twelvett=cmtt12 \font\twelvebf=cmbx12
\font\twelverm=cmr12 \font\twelvei=cmmi12 \font\twelvess=cmss12
\font\twelvesy=cmsy10 scaled \magstep1 \font\twelvesl=cmsl12
\font\twelveex=cmex10 scaled \magstep1 \font\twelveit=cmti12
\font\tenss=cmss10
 
 \font\ninebf=cmbx9
\font\ninerm=cmr9 \font\ninei=cmmi9
\font\ninesy=cmsy9 
\font\eightrm=cmr8
\catcode`@=11
\newskip\ttglue
\newfam\ssfam

\def\fourteenpoint{\def\rm{\fam0\fourteenrm}
\textfont0=\fourteenrm \scriptfont0=\tenrm \scriptscriptfont0=
\sevenrm
\textfont1=\fourteeni \scriptfont1=\teni \scriptscriptfont1=\seveni
\textfont2=\fourteensy \scriptfont2=\tensy \scriptscriptfont2=\sevensy
\textfont3=\fourteenex \scriptfont3=\fourteenex \scriptscriptfont3=
\fourteenex
\def\it{\fam\itfam\fourteenit} \textfont\itfam=\fourteenit
\def\sl{\fam\slfam\fourteensl} \textfont\slfam=\fourteensl
\def\bf{\fam\bffam\fourteenbf} \textfont\bffam=\fourteenbf
\scriptfont\bffam=\tenbf \scriptscriptfont\bffam=\sevenbf
\def\tt{\fam\ttfam\fourteentt} \textfont\ttfam=\fourteentt
\def\ss{\fam\ssfam\fourteenss} \textfont\ssfam=\fourteenss
\tt \ttglue=.5em plus .25em minus .15em
\normalbaselineskip=16pt
\abovedisplayskip=16pt plus 3pt minus 10pt
\belowdisplayskip=16pt plus 3pt minus 10pt
\abovedisplayshortskip=0pt plus 3pt
\belowdisplayshortskip=8pt plus 3pt minus 5pt
\parskip=3pt plus 1.5pt
\setbox\strutbox=\hbox{\vrule height10pt depth4pt width0pt}
\let\sc=\tenrm
\let\big=\fourteenbig \normalbaselines\rm}
\def\fourteenbig#1{{\hbox{$\left#1\vbox to10pt{}\right.\n@space$}}}

\def\twelvepoint{\def\rm{\fam0\twelverm}
\textfont0=\twelverm \scriptfont0=\ninerm \scriptscriptfont0=\sevenrm
\textfont1=\twelvei \scriptfont1=\ninei \scriptscriptfont1=\seveni
\textfont2=\twelvesy \scriptfont2=\ninesy \scriptscriptfont2=\sevensy
\textfont3=\twelveex \scriptfont3=\twelveex \scriptscriptfont3=
\twelveex
\def\it{\fam\itfam\twelveit} \textfont\itfam=\twelveit
\def\sl{\fam\slfam\twelvesl} \textfont\slfam=\twelvesl
\def\bf{\fam\bffam\twelvebf} \textfont\bffam=\twelvebf
\scriptfont\bffam=\ninebf \scriptscriptfont\bffam=\sevenbf
\def\tt{\fam\ttfam\twelvett} \textfont\ttfam=\twelvett
\def\ss{\fam\ssfam\twelvess} \textfont\ssfam=\twelvess
\tt \ttglue=.5em plus .25em minus .15em
\normalbaselineskip=14pt
\abovedisplayskip=14pt plus 3pt minus 10pt
\belowdisplayskip=14pt plus 3pt minus 10pt
\abovedisplayshortskip=0pt plus 3pt
\belowdisplayshortskip=8pt plus 3pt minus 5pt
\parskip=3pt plus 1.5pt
\setbox\strutbox=\hbox{\vrule height10pt depth4pt width0pt}
\let\sc=\ninerm
\let\big=\twelvebig \normalbaselines\rm}
\def\twelvebig#1{{\hbox{$\left#1\vbox to10pt{}\right.\n@space$}}}

\def\tenpoint{\def\rm{\fam0\tenrm}
\textfont0=\tenrm \scriptfont0=\sevenrm \scriptscriptfont0=\fiverm
\textfont1=\teni \scriptfont1=\seveni \scriptscriptfont1=\fivei
\textfont2=\tensy \scriptfont2=\sevensy \scriptscriptfont2=\fivesy
\textfont3=\tenex \scriptfont3=\tenex \scriptscriptfont3=\tenex
\def\it{\fam\itfam\tenit} \textfont\itfam=\tenit
\def\sl{\fam\slfam\tensl} \textfont\slfam=\tensl
\def\bf{\fam\bffam\tenbf} \textfont\bffam=\tenbf
\scriptfont\bffam=\sevenbf \scriptscriptfont\bffam=\fivebf
\def\tt{\fam\ttfam\tentt} \textfont\ttfam=\tentt
\def\ss{\fam\ssfam\tenss} \textfont\ssfam=\tenss
\tt \ttglue=.5em plus .25em minus .15em
\normalbaselineskip=12pt
\abovedisplayskip=12pt plus 3pt minus 9pt
\belowdisplayskip=12pt plus 3pt minus 9pt
\abovedisplayshortskip=0pt plus 3pt
\belowdisplayshortskip=7pt plus 3pt minus 4pt
\parskip=0.0pt plus 1.0pt
\setbox\strutbox=\hbox{\vrule height8.5pt depth3.5pt width0pt}
\let\sc=\eightrm
\let\big=\tenbig \normalbaselines\rm}
\def\tenbig#1{{\hbox{$\left#1\vbox to8.5pt{}\right.\n@space$}}}
\let\rawfootnote=\footnote \def\footnote#1#2{{\rm\parskip=
0pt\rawfootnote{#1}
{#2\hfill\vrule height 0pt depth 6pt width 0pt}}}
\def\tenfoot{\tenpoint\hskip-\parindent\hskip-.1cm}

\twelvepoint
\def\sbullet{\raise.2em\hbox{$\scriptscriptstyle\bullet$}}
\nofirstpagenotwelve
\baselineskip 15pt

\def\ft#1#2{{\textstyle{{#1}\over{#2}}}}

\baselineskip=12pt
\rightline{UR--1259}
\rightline{ER--40685--711}
\rightline{CTP TAMU--31/92}
\rightline{June 1992}

 \vskip 1truecm
\centerline{\bf Self-Duality in 3+3 Dimensions and the KP Equation}
\vskip 1.5truecm
\centerline{Ashok DAS,$^1$\footnote{$^{\star}$}
{Supported in part by the
U.S. Department of Energy, under  grant DE-AC02-76ER-13065.}
Ergin SEZGIN,$^2$\footnote{$^{\dag}$}{Supported in part by the U.S.
National Science Foundation, under grant
PHY-9106593.} and Zurab KHVIENGIA $^2$ }
\vskip
1.5truecm  \item{$^1$}{\it Department of Physics and Astronomy,
University of Rochester, Rochester, \nl NY 14627, USA}
\item{$^2$}{\it Center for
Theoretical Physics, Texas A\&M University, College Station,\nl
TX 77843--4242, USA\/}
\vskip 2.5truecm
\AB\singlespace
\bigskip
   We consider two types of generalized self-duality conditions for
Yang-Mills fields on paracomplex  manifolds of arbitrary dimension.
We then specialize to $3+3$ dimensions and show how one can obtain the KP
equation from these self-duality  conditions on $SL(2,R)$ valued gauge
fields.

      \AE\oneandahalfspace  \np

\noindent
{\bf 1. Introduction}
\bigskip
 There are many physical systems that are described by nonlinear,
 Hamiltonian
equations which are integrable, i.e., they can be solved exactly [1,2].
These systems
possess many interesting properties which make them quite attractive.
In recent
years, a growing number of connections have been found between such
systems and some
of the models of theoretical high- energy physics. For example, it is known
that the
various conformal algebras [3,4] ($W_N$ algebras etc.) arise as
 the Hamiltonian structures (Poisson brackets) of the integrable systems.
 In fact,
the first Hamiltonian structure [5] of the 2+1 dimensional integrable
system, the KP
hierarchy [6], can be identified with the $W_\infty$ algebra [7] which
has been a subject of much recent work in diverse areas, e.g. in 2D gravity
[8].

 In a separate development, it was found that the 1+1 dimensional
 integrable systems can be obtained from the self-duality
conditions imposed on the  Yang-Mills potentials in four and
higher dimensions [9]. Thus, for example, it is known that the  KdV equation,
the nonlinear Schr\"odinger equation, etc., can be obtained from the
self-duality condition on the SL(2,R) valued Yang-Mills potentials in four
dimensions with (2,2) signature upon appropriate reduction [10]. Similarly,
the  generalized KdV equations (namely, Boussinesq equation etc.) can also be
obtained from the  self-duality condition for higher SL(N,R) groups [11]. It is
also known that the KP  hierarchy leads to most of the 1+1 dimensional
integrable models upon appropriate  reduction. It is, therefore, quite natural
to expect that the KP equation can also be obtained  from a
self-duality condition of the Yang-Mills potential in higher dimensions.
Various attempts in this direction, however, have been unsuccesful and the
reason  for the failure is normally attributed to the nonlocal nature of the
scalar Lax  operator for the KP equation.

The derivation of various 1+1 or 2+1 dimensional integrable systems from
self-duality
conditions in higher dimensions provides a unified picture for seemingly
unrelated vast class of integrable systems. As a byproduct, it may in
principle give  rise to new integrable systems. Moreover, the symmetries of
all these integrable  systems can now be understood in terms of gauge
transformations in higher dimensions.  This not only provides a way to
understand these symmetries, but also provides  a way to search for new
hidden symmetries, if any.

In this letter, inspired by the successful embedding of the KdV equation in
the self-duality condition on the Yang-Mills potential in 2+2 dimensions,
which  we shall outline in Section 2, we analyze the self-duality
conditions for Yang-Mills potentials in 3+3 dimensions and show how the
KP equation follows from such a condition on SL(2,R) valued Yang-Mills
potentials. In fact, in Section 3, we shall describe two types of
generalized self-duality  conditions on paracomplex manifolds [12] of
arbitrary dimension, and then specialize to $3+3$ dimensions.
In Section 4, we shall derive the KP equation from
these self-duality equations. In the last section, we shall comment on open
problems.
\bigskip
\noindent{\bf 2. Self-Duality in 2+2 Dimensions and the KdV Equation}
\bigskip
In this section we shall outline  the embedding of KdV
equation in self-duality condition  in 2+2 dimensions [10,11]. Consider the
SO(2,2) invariant line element
$$
  ds^2=2dxdy+2dzdt.  \eqno(1)
$$
The metric $\eta_{\mu\nu}$ is defined by $ds^2=
\eta_{\mu\nu}dx^\mu dx^\nu $, where
$x^\mu=(x,y,z,t)$, and evidently has the signature $(+,-,+,-)$. The
self-duality condition is
$$
F_{\mu\nu}=\ft12 \epsilon_{\mu\nu\rho\sigma} F^{\rho\sigma},  \eqno(2)
$$
where the indices of the field strength $F_{\mu\nu}$ for the Yang-Mills
potential
$A_\mu$ are raised and lowered by the metric $\eta_{\mu\nu}$. Eq. (2) is
equivalent
to the following three equations (with $\epsilon_{xyzt}=-1$)
$$
\eqalignno{
            F_{tx}&=0, &(3) \cr
            F_{yz} &=0, &(4)\cr
            F_{tz}+F_{xy} &=0. &(5)  \cr}
$$
Let us identify $t,x$ as the coordinates of the 1+1 dimensional
spacetime, and use the notation $A_t=H,\ A_x=Q,\ A_y=P,\ A_z=-B$.
Imposing
the conditions
$$
   \partial_z =0, \quad\quad \partial_y-\partial_x =0,    \eqno(6)
$$
and taking  $B$ to be constant, the self-duality equations (3-5)  reduce to
$$
\eqalignno{
     &[\partial_t-H, \partial_x-Q] =0,  &(7)\cr
     &[P,B]=0,  &(8)\cr
     &[H,B]=[\partial_x-Q, \partial_x -P].  &(9) \cr}
$$
We need to make further ansatz for the gauge
potentials in order to derive particular integrable systems from (7-9). Let
the gauge
potentials take their value in SL(2,R), and consider the following ansatz
$$
  B=\left(\matrix{0&0\cr -1&0\cr}\right),\quad\quad\quad
  Q=\left(\matrix{\lambda&1\cr -u&-\lambda}\right), \eqno(10)
$$
where $u$ is an arbitrary function of $t,x+y$ and $\lambda$ is an arbitrary
constant.
In terms of the matrices
$$
\sigma_+=\left(\matrix{0&1\cr 0&0\cr}\right),\qquad
\sigma_-=\left(\matrix{0&0\cr 1&0\cr}\right),\qquad
\sigma_3=\left(\matrix{1&0\cr 0&-1\cr}\right),\qquad  \eqno(11)
$$
which obey the SL(2,R) algebra
$$
[\sigma_+, \sigma_-]=\sigma_3,\qquad\qquad [\sigma_3,\sigma_{\pm}]=
\pm 2\sigma_{\pm},
\eqno(12)
$$
expanding the functions $H=H_3 \sigma_3+H_-\sigma_+ +H_+\sigma_- $
and
$P=P_3 \sigma_3+P_-\sigma_+ +P_+\sigma_- $, first, from (8) one
determines that
$$
 P_- =0,\quad\quad P_3 =0. \eqno(13)
$$
Next, from (9) one learns that
$$
     H_- =-P_+, \quad\quad H_3 =-\ft12 (u+P_+)'-\lambda P_+\ .   \eqno(14)
$$
The functions $H_+$ and $P_+$ are still undetermined. Using (10) and (14)
in the
flatness condition (7), on the other hand, we obtain the equations
$$
\eqalignno{
&(u+2P_+)' =0, &(15)\cr
     &H_+ =uP_+ -\lambda P_+' -\ft12(u+ P_+)'' ,   &(16)\cr
&{\dot u} =\ft12(u+ P_+)''' +(u-P_+)u'\ +2\lambda^2 P_+'. &(17)\cr}
$$
{}From (15), setting the integration constant equal to zero we obtain
$P_+=-\ft12 u$, and consequently $H$ and $P$ are now completely
determined in terms of
a single function $u(t,x+y)$ and (17) now reads
$$
{\dot u}=\ft14 u''' +\ft32 u u'-\lambda^2 u'. \eqno(17a)
$$
Shifting $u$ by a constant as $u\rightarrow u+\ft23 \lambda^2$, this
equation becomes
identical to the usual KdV  equation:
$$
{\dot u}=\ft14 u''' +\ft32 u u', \eqno(17b)
$$
where $t$ and $(x+y)\equiv x_+$, can be viewed as the time and space
coordinates of a 1+1 dimensional spacetime and $'$ now denotes
$\partial/\partial x_+$.

In summary, with the reduction conditions (6) and choices
(10) for the SL(2,R) potentials, the self-duality condition (2) leads to the
KdV
equation. It turns out that different choices for the connection $Q$ lead to
different
integrable systems in 1+1 dimensions. However, a hierarchy of integrable
systems such
as the KdV hierarchy does not seem to follow from the self-duality
equations. In the
case of KdV hierarchy,  the flatness condition (7) alone can accommodate it
by suitable choices of $H$ and $Q$, but those choices would be incompatible
with the remaining equations (8) and (9)  except for the case of KdV
equation.

As mentioned earlier, it is desirable to obtain also the KP equation from a
self-duality condition in a higher dimension. Interestingly enough, if we
don't shift away the $\lambda$ term in (17a), but instead take the
$\partial_x$  derivative
of both sides and use the constraint $ \partial_x u=\partial_y u$ in the
$\lambda$ dependent term, we obtain the KP equation
described in Sec. 4 (after rescaling $u \rightarrow 2u$ and setting
$\lambda={\sqrt 3}/2$\footnote{$^\dagger$}{\tenfoot This procedure yields
the KP  equation (37). By
choosing the integration constant arising from (15) appropriately, we
can obtain
also the KP equation (36).}). Note, however, that the constraint  $\partial_x
u=\partial_y u$ is imposed by hand. Furthermore, it  implies a 1+1
dimensional structure for the KP equation.   In Sec. 4, we shall
 show how such a constraint as well as the KP equation arises from certain
types of  self-duality conditions in 3+3 dimensions.  We shall than
consider an alternative choice of gauge potentials which will lead to the
KP equation, without the constraint $\partial_x u=\partial_y u$.
We now turn to the description of these  self-duality conditions.
\bigskip
\noindent{\bf 3. Self-Duality Conditions in 3+3 Dimensions}
\bigskip
Linear relationships among the components
of the field strength $F_{\mu\nu}$ in $d$-dimensional Euclidean space of
the following type
$$
\lambda F_{\mu\nu}=\ft12 T_{\mu\nu\rho\sigma} F^{\rho\sigma}, \eqno(19)
$$
where the tensor $T_{\mu\nu\rho\sigma}$ is totally antisymmetric,   and
$\lambda$ is a constant, have been considered in [13]. In four dimensions one
can choose the T-tensor to be the Levi-Civita symbol which is $SO(4)$
invariant. However,   in higher
than four dimensions an $SO(d)$ invariant T-tensor is not available.
Nonetheless, in some cases one can find a T-tensor which is invariant under
 a subgroup of $SO(d)$. Apart from this, the iteration of (19), of course, puts
a severe restriction on the T-tensor. A number of examples in $d\le 8$ have
been  provided in [13].  Further examples are given and the issue of which
self-duality conditions can be deduced from the integrability condition of a
linear equation was studied in [14].

Here we shall consider self-duality condition of the form (19) in a six
dimensional space of signature (3,3). This is the simplest case beyond four
dimensions in which we
may seek the analogs of the (2,2) signature self-duality equtions of the
previous  section. We
shall also consider another type of self-duality equation in which the
T-tensor is not required to be totally antisymmetric.

Both of these self-duality conditions will be formulated on paracomplex
manifolds [12].  A paracomplex manifold M is a manifold  which admits a
(1,1) type tensor $J_\mu^\nu$ which satisfies the condition
$$
     J_\mu^{\ \nu} J_\nu^{\ \rho} =\delta_\mu^\rho.  \eqno(20)
$$
If M admits a metric $g_{\mu\nu}$ which satisfies the condition
$$
J_\mu^{\ \rho} g_{\rho\nu}=-J_\nu^{\ \rho} g_{\rho\mu}, \eqno(21)
$$
then $g$ will be called parahermitian metric, and $M$  a parahermitian
manifold.

Let M be a parahermitian manifold of dimension $d$.  Our first self-duality
 condition on the field strength $F_{\mu\nu}$ is
$$
Case\ 1:\quad \quad\quad  F_{\mu\nu}=\ft32
J_{[\mu\nu}J_{\rho\sigma]}F^{\rho\sigma}, \eqno(22)
$$
where $J_{\mu\nu}$ is the antisymmetric tensor defined by
$J_{\mu\nu}=J_\mu^{\ \rho} g_{\rho\nu}$ and the antisymmetrizations are
 with unit
strength, e.g. $J_{[\mu\nu}J_{\rho\sigma]}=
\ft13(J_{\mu\nu}J_{\rho\sigma}
-J_{\mu\rho}J_{\nu\sigma}+J_{\mu\sigma}J_{\nu\rho})$. Iterating (22)
one finds that
it is consistent by itself only in $d=4$, while in dimensions other than
 four  it is equivalent to the following equations
$$
    F_{\mu\nu}=-J_\mu^{\ \rho}J_\nu^{\ \sigma} F_{\rho\sigma}\qquad
    {\rm and} \qquad J^{\mu\nu} F_{\mu\nu} =0.   \eqno(23)
$$

Let us now specialize to a flat six dimensional space of signature (3,3).
We can choose the metric to be
$$
      g_{\mu\nu}=\oneone_3 \otimes \sigma_1. \eqno(24)
$$
In this coordinate system we can put the paracomplex structure
in the   following form \footnote{$^\dagger$}{\tenfoot We could have chosen
the metric to take the form
(25) in which case the paracomplex structure must take the form (24).}
$$
 J_\mu{}^\nu=\oneone_3 \otimes \sigma_3.  \eqno(25)
$$
With these choices, the equations (22) and (23) read
$$
\eqalignno{
&F_{13}=0,\qquad F_{16}=0, \qquad F_{36}=0,      &(26)\cr
&F_{24}=0,\qquad F_{25}=0, \qquad F_{45}=0,     &(27)\cr
&F_{12}+F_{34}+F_{65} =0.     &(28) \cr}
$$
These equations are rather similar to a set given in [13] in the case of
Euclidean
signature. In that case the invariance group of the self duality equations is
$SU(3)\times U(1)$, while in our case the invariance group is
$SL(3,R)\times SO(1,1)$.  In both cases, it is  not known if these
equations follow from the integrability conditions of a set of linear
equations.

We now turn to the description of another type of self-duality condition
which we
can interpret as the integrability condition for a linear equation in a
paracomplex
manifold, $M$, in general. We shall subsequently specialize to the flat 3+3
dimensional space. The self-duality condition amounts to stating that
$F_{\mu\nu}$ is
pure, in the sense that given the projectors
$P_{\pm\mu}^{\ \ \ \nu}=\ft12(\delta_\mu^\nu \pm J_\mu^{\ \nu})$, it
satisfies, for
example, $P_{+\mu}^{\ \ \ \rho} P_{+\mu}^{\ \ \ \sigma} F_{\rho\sigma}=0$.
This can be written as
$$
  Case\ 2 :\quad\quad\quad   F_{\mu\nu}=-\big(J_\mu^{\ \rho}
  J_\nu^{\ \sigma}
    +2J_\mu^{\ \rho}\delta_\nu^\sigma\big) F_{\rho\sigma}.   \eqno(29)
$$
The iteration of this condition gives a consistent result, and no new
conditions
arise. If we view this condition in the form (19), the T-tensor here is not
totally antisymmetric, and hence this falls outside the class of self-duality
equations considered in [13]. Note furthermore, that the metric does not
occur in (29),  and
therefore the symmetry group of this equation is huge, namely, all the
paraholomorphic transformations of $M$, i.e. those coordinate
transformations which preserve the form of the paracomplex structure $J$.

The self-duality condition (29) can be easily seen to follow as the
integrability condition of the following linear equation
$$
    (\delta_\mu^\nu + J_\mu^{\ \nu})D_\nu \psi=0, \eqno(30)
$$
where $\psi$ is an arbitrary function in some representation of the
Yang-Mills gauge
group, and the gauge covariant derivative is $D_\mu=\partial_\mu-A_\mu$.
If we
specialize to a space of signature (3,3), using the fact that it is possible to
have a Majorana-Weyl spinor in this space, and some $\gamma$ matrix
identities special to this case, we can show that the linear equation (30) is
equivalent to the following
$$
   \gamma^\mu\eta D_\mu \psi=0, \eqno(31)
$$
where $\eta$ is a commuting Majorana-Weyl spinor, and $\gamma^\mu$ are
the Dirac $\gamma$-matrices in 3+3 dimensions obeying the Clifford
algebra  $\{\gamma_\mu,\gamma_\nu\}=2g_{\mu\nu}$. To show this,
consider  another
commuting Majorana-Weyl spinor of opposite chirality $\chi$ which is
normalized such
that  ${\bar \chi} \eta =1$. Multiplying (31) from the left with ${\bar \chi}
\gamma_\nu$ we obtain
$$
{\bar \chi} \gamma_\nu\gamma^\mu\eta D_\mu \psi=0\ \Rightarrow \
\big(\delta_\nu^\mu+
  {\bar \chi}\gamma_\nu^{\ \mu}\eta \big)D_\mu \psi=0.  \eqno(32)
$$
It remains to show that
$$
    {\bar \chi}\gamma_\nu^{\ \mu}\eta =J_\nu^{\ \mu}.  \eqno(33)
$$
Indeded, by Fierz rearrangement we find that
$$
\eqalign{
J_\nu^{\ \mu}J_\mu^{\ \rho}&={\bar \chi}\gamma_\nu^{\ \mu}\eta
  {\bar \chi}\gamma_\mu^{\ \rho}\eta  \cr
  &=\ft54\delta_\nu^\mu-\ft18 J^2-J_\nu^{\ \mu}J_\mu^{\ \rho}, \cr}
  \eqno(34)
$$
where $J^2=J_\nu^{\ \mu}J_\mu^{\ \nu}$. Taking the trace of this equation
one finds $J^2=6$, and substituting this back to (34) one finds that indeed
$J_\nu^{\ \mu}J_\mu^{\ \rho}=\delta_\nu^\rho$. (An analogous calculation
can be found
in [15] for the case of a complex structure on a six dimensional manifold of
Euclidean signature).

It is worthwhile mentioning that the vector $u^\mu=
{\bar \lambda}\gamma^\mu\eta$
where $\lambda$ is a commuting Majorana-Weyl spinor of the same
chirality as $\eta$,
is a null vector, i.e. $u^\mu u_\mu=0$. Let there be $N$ such vectors,
$u^\mu_{(r)},\
r=1,...,N$. Then, the linear equation (31) can also be written as
$u^\mu_{(r)}D_\mu \psi=0$.  Note also that a commuting Majorana-Weyl
spinor $\eta$ in 3+3 dimensions is pure, in the sense that
${\bar \eta}\eta=0$.

Finally, using (24) and (25) we find that the self-duality condition (29)
reduces to the following three equations
$$
  F_{13}=0,\qquad F_{16}=0,\qquad F_{36}=0. \eqno(35)
$$
\bigskip
\noindent{\bf 4. The KP Equation}
\bigskip
The KP equation is a dynamical equation in 2+1 dimension and has the form
$$
\partial_x\left({\dot u}-\ft14 u_{xxx} -3uu_x\right)= \ft34 u_{yy},
\eqno(36)
$$
where $u(x,y,t)$ is the dynamical variable, and subscripts $x$ and $y$
denote partial differentiations with respect to $x$ and $y$,  respectively.
For a medium with an opposite  dispersive behaviour, the equation takes the
form
$$
\partial_x\left({\dot u}-\ft14 u_{xxx} -3uu_x\right)= -\ft34 u_{yy},
\eqno(37)
$$

We shall first consider the embedding of the KP equation in the
self-duality equtions
(26-28). To this end, we identify the coordinates $x^1, x^3, x^6$ with the
coordinates  $t,x,y$ of a 2+1 dimensional spacetime and let
$$
\eqalign{
   &A_1=H,\qquad A_3=Q,\qquad A_6={\tilde Q}, \cr
   &A_2=-B,\qquad A_4=P,\qquad A_5={\tilde B}. \cr}  \eqno(38)
$$
Next, we impose the reduction conditions
$$
\partial_2=0,\quad \partial_4-\partial_x=0,\quad \partial_5=0,\eqno(39)
$$
and take $B$ and ${\tilde B}$ to be constant.  With these
conditions, the self-duality equations (26-28) reduce to
$$
\eqalignno{
     &[\partial_t-H, \partial_x-Q] =0,  &(40)\cr
     &[\partial_t-H, \partial_y-{\tilde Q}] =0,  &(41)\cr
     &[\partial_x-Q, \partial_y-{\tilde Q}]=0,  &(42)\cr
     &[P,B]=0, &(43)\cr
     &[P,{\tilde B}]=0, &(44)\cr
     &[B,{\tilde B}]=0, &(45)\cr
     &[H,B]=[\partial_x-Q, \partial_x-P]-[{\tilde Q}, {\tilde B}].  &(46)\cr}
$$
Finally we need to make ansatz for the gauge fields $Q,\ {\tilde Q}$ and
$B$. As before, we take the gauge group to be SL(2,R). Next, for $Q$
and $B$ we again take the ansatz (10), and  in addition we take
${\tilde B}$ to be proportional to $B$ and  ${\tilde Q}$  proportional to $Q$.
More specifically, we make the following ansatz
$$
  B=-\sigma_-,\qquad Q=\lambda \sigma_3-u\sigma_-+\sigma_+, \qquad
  {\tilde B}=\alpha \sigma_-, \qquad {\tilde Q}=\beta Q, \eqno(47)
$$
where $\alpha, \beta$ and $\lambda $ are arbitrary constants. The
consequences of these ansatz are as follows. From (42) we learn that
$$
  u_y=\beta u_x. \eqno(48)
  $$
Consequently, (41) reduces to (40). Next, from (43-44) we see that
$$
 P_- =0,\quad\quad P_3 =0. \eqno(49)
$$
As in the case of KdV, the  constraint (46) determines  $H_-$ and $ H_3$:
$$
\eqalign{
  H_- &=-(\alpha\beta +P_+), \cr
 H_3 &=-\ft12 (u+P_+)'-\lambda (\alpha\beta+P_+)\cr}\eqno(50).
$$
Finally, from (40) we find that
$$
\eqalignno{
  &(u+2P_+)' =0,   &(51) \cr
  &H_+ =u(\alpha\beta+ P_+) -\lambda P_+' -\ft12(u+ P_+)'' ,   &(52)\cr
 &{\dot u} =\ft12(u+ P_+)''' +(u-\alpha\beta-P_+)u'\ +2\lambda^2 P_+'.
 &(53)\cr}
$$
Note that (50) and (51-53) are related to (14-17) by shifting
$P_+\rightarrow
P_++\alpha\beta$ in the latter set of equations. As before, choosing the
integration
constant in (51) to be zero, we obtain $P_+=-\ft12 u$, and therefore all the
components of the gauge potential are now determined, and (53) reduces to
$$
{\dot u} -\ft14 u'''-\ft32 uu'=-(\lambda^2+\alpha\beta) u'\ .  \eqno(54)
$$
Differentiating this eqution with respect to $x$ and then using the
constraint (48), we obtain
$$
\partial_x\left({\dot u}-\ft14 u_{xxx}-\ft32 uu_x\right)=
-\left({\lambda^2+\alpha\beta\over \beta^2}\right) u_{yy}. \eqno(55)
$$
Rescaling $u\rightarrow 2u$ and setting ${\lambda^2+\alpha\beta\over
\beta^2}=\pm \ft34$, yields precisely the KP equations (36) and (37).

Given the ansatz (47) for the gauge potentials $Q$ and ${\tilde Q}$, the
results
(50-52) for the Hamiltonian, it is clear that the vanishing curvature
equation in
2+1 dimensions for an SL(2,R) valued gauge potential yields the KP
equation, as was
shown in [16]. What we have found here is that the reduction scheme
summarized in (39) and (47) is actually sufficient to determine the
Hamiltonian and that the resulting dynamical equations are the KP
equations. In other words, tuning the gauge potentials in spatial
directions plus the self-duality condition determine the Hamiltonian and
the dynamical field equations. If we were given only the vanishing
curvature equations, we would have to make choices for the gauge potentials
in spatial directions {\it and} the Hamiltonian to obtain the desired
dynamical  equations.

In fact, this is the situation we encounter in the case of the second-kind of
self-duality condition (35), which happens to be the vanishing curvature
condition; i.e. the ansatz (47) for the gauge potentials $Q$ and ${\tilde Q}$,
the results  (50) and (51) for the Hamiltonian, which should themselves now
be viewed as  ansatz, of course, provide the embedding of the KP equation in
the self-duality  condition (35).

Going back to the ansatz (47),  we saw that it gave rise to the constraint
equation (48). As mentioned before, while this constraint is derived from the
self-duality equation, it  implies a $1+1$ dimensional structure for the KP
equation. We can, in fact, derive the KP equation from self-duality
conditions (40-46) without such a constraint. For example, let us consider
the following alternative  choice of the potentials
$$
 B=\alpha \sigma_-,\qquad  {\tilde B}=\beta \sigma_-,
\qquad Q=P=u\sigma_-, \qquad  {\tilde Q}={4\lambda\over
3}\partial_y^{-1}\phi_x \sigma_-, \eqno(56)
$$
where $u$ and $\phi$ are the dynamical variables and $\alpha,\beta,\lambda$ are
arbitrary constants.  Note that
$\partial_y^{-1}$  corresponds to an indefinite integration over the
$y$--coordinate.  Given this ansatz, we see that the self-duality equations
(43-45) are trivially satisfied,  and that (42) yields the result
$$
\phi_{xx}={3\over 4\lambda} u_{yy},    \eqno(57)
$$
and (46) is satisfied with
$$
 H_- =0,\quad\quad H_3 =0. \eqno(58)
$$
Using (57), one finds that (41) is satisfied provided that (40) is satisfied.
This is the dynamical equation
$$
{\dot u} = \partial_x H_+ .  \eqno(59)
$$
It is clear that, in this case, we need to make an appropriate ansatz for
$H_+$ in order to obtain the KP equation. This ansatz is
$$
H_+=\left(\ft32 u^2 +\ft14 u_{xx} +\lambda \phi\right) \sigma_-. \eqno(60)
$$
Indeed, substituting (60) into (59) and using (57) we obtain the KP equation
(36).  This derivation, in fact, is very similar in spirit to the original
derivation of the KP equation [6]. We note, in this case, that the derivation
would go through even if we assume the gauge group to be $U(1)$.

 \bigskip
\noindent{\bf 5. Conclusions}
\bigskip
Contrary to the common belief that the KP equation would not result from
 the self-duality condition on the Yang-Mills potentials, we have explicitly
 derived these equations from self-duality conditions in 3+3 dimensions. The
two types of self-duality conditions considered in this paper, we believe,
are interesting to study further in their  own right.

An interesting open question that remains open is the derivation of the KdV
and/or KP
{\it hierarchy} from the self-duality conditions. The presence of extra
internal dimensions studied here, may indeed have relevance to this
question.  Another
interesting open problem is to determine if choices less
stringent than those in (47) on the gauge potentials would lead to new
integrable
systems in 2+1 or higher dimensions, and in any event to determine the
class of
integrable systems for which our self-duality conditions provide a unified
framework.
\bigskip\bigskip
\centerline{\bf ACKNOWLEDGEMENTS}
We thank HoSeong La and Chris Pope for useful discussions. We are grateful to
Steve Fulling for bringing to our attention the references on paracomplex
numbers.
\vfill\eject \bigskip
\centerline{\bf REFERENCES}
\bigskip
\singlespace
\item{1.} L.D. Faddeev and L.A. Takhtajan, {\it Hamiltonian methods in the
theory of
solitons} (Springer, Berlin, 1987).
\item{2.} A. Das, {\it Integrable Models} (World Scientific, 1989).
\item{3.} F. Magri, J. Math. Phys. {\bf 19} (1978) 1156.
\item{} J.L. Gervais, Phys. Lett. {\bf 160B} (1985) 277.
\item{4.} V. Drinfeld and V. Sokolov, J. Sov. Math. {\bf 30} (1985) 1975.
\item{} A. Bilal and J.L. Gervais, Phys. Lett. {\bf 206B} (1988) 412.
\item{} I. Bakas, Nucl. Phys. {\bf B302} (1988) 189.
\item{} A. Das and S. Roy, Int. J. Modern Phys. {\bf 6} (1991) 1429.
\item{5.}Y. Watanabe, Ann. Math. Pure Appl. {\bf 136} (1983) 77.
\item{} K. Yamagishi, Phys. Lett. {\bf 259B} (1991) 436.
\item{} F. Yu and Y.S. Wu, Phys. Lett. {\bf 263B} (1991) 220.
\item{} A. Das, W.J. Huang and S. Panda, Phys. Lett. {\bf 271B} (1991) 109.
\item{6.} B.B. Kadomtsev and V.I. Petviashvili, Sov. Phys. Dokl. {\bf 15}
(1971) 539.
\item{} E. Date, M. Kashiwara, M. Jimbo and T. Miwa, in {\it Nonlinear
integrable systems}, eds. M. Jimbo and T. Miwa (World Scientific, Singapore,
1983).
\item{7.} C.N. Pope, L.J. Romans and X. Shen, Phys. Lett. {\bf 236B}
(1990)  173;\ Nucl. Phys. {\bf B339} (1990) 191;\ Phys. Lett. {\bf 242B}
(1990) 401.
\item{8.} A.M. Polyakov, Mod. Phys. Lett. {\bf A2} (1987) 893.
\item{} V.Z. Knizhnik, A.M. Polyakov and A.B. Zamolodhikov, Mod. Phys. Lett.
{\bf A3} (1988) 819.
\item{9.} R.S. Ward, Nucl. Phys. {\bf B236} (1984) 381;\ Philos.
Trans. R. Soc. {\bf A315} (1985) 451;\ in {it Field Theory, Quantum Gravity
and Strings},  Vol. 2, p.106, eds. H.J. DeVega and N. Sanchez.
\item{10.} L.J. Mason and G.A.J. Sparling, Phys. Lett. {\bf 137A} (1989) 29.
\item{11.} I. Bakas and D.A. Depireux, Int. J. Mod. Phys. {\bf A7} (1992) 1767.
\item{12.} P. Libermann, C.R. Acad. Sci. Paris {\bf 234} (1952) 2517; \ Ann.
Math. Pura Appl. {\bf 36} (1954) 27.
\item{} M. Frechet, Compositio Math. {\bf 12} (1954) 81.
\item{} S. Kaneyuki, Tokyo J. Math. {\bf 8} (1985) 473;\  Japan. J. Math.
{\bf 13} (1987) 333.
\item{13.}  E. Corrigan, C. Devchand, D.B. Fairlie and J. Nuyts, Nucl. Phys.
{\bf B214} (1983) 452.
\item{14.} R.S. Ward, Nucl. Phys. {\bf B236} (1984) 381.
\item{15.} C.N. Pope and P. van Nieuwenhuizen, Commun. Math. Phys. {\bf 122}
 (1989) 281.
 \item{16.} J. Barcelos-Neto, A. Das, S. Panda and S. Roy, Phys. Lett.
{\bf 282B} (1992) 365.

\end